\documentstyle[aps,preprint,prb]{revtex}
\begin{document}
\tightenlines
\def\vk{\vec k} 
\def\br{{\bf r}}
\title{\bf A New Interpretation of Flux Quantization }
\author{Mi-Ae Park}
\address{Department of Physics,  University of Puerto Rico \\
 Humacao, PR 00791}
\author{Yong-Jihn Kim}
\address{Department of Physics, University of Puerto Rico \\
 Mayaguez, PR 00681 }
\maketitle
\begin{abstract}
We study the effect of Aharonov-Bohm flux on the superconducting state in 
metallic cylinders. Although Byers and Yang attributed flux quantization to 
the flux-dependent minimum of kinetic energies of the Cooper pairs, it is shown
that kinetic energies do not produce any discernible oscillations in the free 
energy of the superconducting state (relative to that of normal state) 
as a function of the flux. This result is indeed anticipated by the 
observation of persistent current in normal metal rings at low temperature. 
Instead, we have found that pairing interaction depends on the flux, 
leading to flux quantization.
When the flux $(\Phi$) is given by $\Phi=n\times hc/2e$ (with integer n), 
the pairing interaction and the free energy become unchanged (even n) or
almost unchanged (odd n), due to degenerate-state pairing 
resulting from the energy level crossing. As a result, 
flux quantization and Little-Parks oscillations follow. 
\end{abstract}
\vskip 1pc

\newpage
\section{\bf Introduction} 

Flux quantization is one of the most fascinating properties of superconducting 
states. This phenomenon was found by Deaver and Fairbank$^{1}$ and Doll and 
N\"abauer$^{2}$ in 1961. They observed that the magnetic flux trapped in a 
superconducting cylinder is quantized in units of hc/2e.
Recently, half-integer flux quantization has attracted  
much attention in connection with the pairing symmetry in high Tc cuprates.$^{3}$

Byers and Yang$^{4}$ proposed that flux quantization follows because the free
energy of the superconducting state is periodic in the unit of the flux, hc/2e,
if electrons are paired. In fact, they assumed that i) the free energy, $F_{N}$, of the 
normal state is independent of the flux, i.e., $F_{N}(\Phi)=\rm{constant}$, and ii) the pairing energy is constant
as a function of the flux.$^{5}$ 
Consequently, they ascribed flux quantization to the minimum of the kinetic 
energy of the superconducting state at $\Phi/(hc/2e)=\rm{integer}$. 
Based on the theory of Byers and Yang,
Little and Parks$^{6}$ pointed out that the 
transition temperature $T_{c}$ is also a periodic function of the enclosed 
flux $\Phi$, which is called Little-Parks oscillations.

However, recent discovery of persistent current in normal metal 
rings at low temperature casts a doubt on the first assumption, that the free 
energy of the normal state is essentially independent of the flux.$^{7-9}$
Since the electronic states of normal metal rings are periodic in 
flux $\Phi$, with period hc/e, the persistent current is given by 
$I(\Phi)=-c\times \partial F_{N}(\Phi)/\partial \Phi$. Therefore, the
first assumption is not compatible with the experimental observation of the 
persistent current in normal metal rings.  

In this paper we report that 
flux quantization and the Little-Parks oscillations are caused by the flux 
dependence of the pairing energy. In addition, we show that
the flux dependence of the kinetic energy
does not lead to any significant oscillations in the transition temperature 
$T_{c}$ and the energy gap parameter $\Delta_{0}$ for superconducting 
cylinders.
As the flux increases, the pairing matrix element is reduced since we must pair
single-particle states which are not degenerate and have different density
distributions.$^{10,11}$
When the flux $(\Phi$) is equal to $\Phi=n\times hc/2e$ (with integer n), 
the pairing interaction and the free energy of the superconductor remain unchanged (even n) or almost
unchanged (odd n) from their values with zero flux, since we can pair degenerate partners,
with the same density distribution, due to the energy level crossing. Accordingly, 
we find flux quantization and Little-Parks oscillations. 
It is interesting to note that Schwartz and Cooper$^{12}$ obtained a similar
result for {\sl generalized pairing} with different angular momentum in 
superconducting cylinders.

\section{\bf Theory of Byers and Yang} 

To study the effect of the Aharonov-Bohm flux $\Phi$ on the superconducting 
states in cylinders, Byers and Yang$^{4}$ considered the variation of the energy
levels E of a single particle in the presence of the flux:
\begin{equation}
E={1\over 2m}[p_{r}^{2}+p_{z}^{2}+{\hbar^{2}\over r^{2}}(m+{e\over hc}\Phi)^{2}].
\end{equation}
Here $p_{r}$ and $p_{z}$ denote momenta in the radial and z directions.
Employing the BCS pairing between states m and -m, they noted that the average 
(kinetic) energy for the states m and -m, near 
$\Phi=+0$, increases with $\Phi$ like
\begin{equation}
\rm{ constant} + {\hbar^{2}\over 2mr^{2}}({e\Phi\over hc})^{2}.
\end{equation}
Furthermore, at $2e\Phi/hc=1$, since pairing between 
\begin{equation}
m+{e\over hc}\Phi=m+{1\over 2}={1\over 2}\ \rm{and} -{1\over 2}, {3\over2}\ \rm{and} -{3\over 2},\ \rm{etc.},
\end{equation}
is preferred,  the kinetic energy per particle remains the same as for the case 
$\Phi=0$. 
Then the additional energy for each of these pairs, in the neighborhood of 
$2e\Phi/hc$, increases with $\Phi$ in a parabolic manner:
\begin{equation}
2\times({\hbar^{2}\over 2mr^{2}})[({e\Phi\over hc})-{1\over 2}]^{2}.
\end{equation}

Consequently, the kinetic energy of the Cooper pairs has the minimum
at $\Phi=n\times hc/2e$ (with integer n). 
Based upon this observation, Byers and Yang$^{4}$ asserted that 
flux quantization is due to the flux-dependent minimum of the kinetic energy 
of the Cooper pairs.
On the other hand, the corresponding BCS wavefunctions at $\Phi=n\times hc/2e$  
are given by$^{13}$
\begin{equation}
\prod(u_{\nu}+v_{\nu}a^{+}_{\nu+m\uparrow}a^{+}_{-\nu+m\downarrow})|0>;\ (n\ \rm{even})
\end{equation}
\begin{equation}
\prod(u_{\nu+1/2}+v_{\nu+1/2}a^{+}_{\nu+m+1\uparrow}a^{+}_{-\nu+m\downarrow})|0>;\ (n\ \rm{odd})
\end{equation}
with $n=2m\ (n=2m+1)$ for even (odd) values of n.
The operator $a^{+}_{\nu+m\uparrow}$ creates an electron in the state 
$(\nu+m\uparrow)$. 

However, the kinetic energy of the normal electrons will show the same flux
dependence, as confirmed
by the observation of the persistent current in normal metal rings.$^{7-9}$
Therefore, we expect that this flux dependence will be canceled in the 
superconducting state.
More precisely, using the BCS theory, we show that the variation of the kinetic 
energy with the flux does not
lead to any oscillations of the condensation energy (i.e., $\Delta_{0}$) and
the free energy (i.e., $T_{c}$) of superconducting cylinders.

Assuming a constant pairing matrix element V,$^{4,5}$ one finds the BCS gap 
equation at $T=0K$ and the $T_{c}$ equation:$^{14-16}$
\begin{equation}
\Delta_{m}=-V\sum_{m'}{\Delta_{m'}\over 2\sqrt{({\epsilon_{m'-\mu}+\epsilon_{-m'-\mu}\over 2})^{2}+\Delta_{m'}^{2}}}
\end{equation}
\begin{equation}
1=-V\sum_{m}{tanh{\epsilon_{m-\mu}\over 2T_{c}}+tanh{\epsilon_{-m-\mu}\over 2T_{c}}\over 2(\epsilon_{m-\mu}+\epsilon_{-m-\mu})}
\end{equation}
where $\epsilon_{m-\mu}$ is the one-particle energy (measured from the Fermi 
energy $E_{F}$) with $\mu=e\Phi/hc$.
Contrary to Byers and Yang,  it is clear that the kinetic energy  
does not give rise to any significant oscillations in the gap parameter, 
$\Delta_{m}=\Delta_{0}$, since its effect is only to shift the BCS 
cutoff  region from $-\omega_{D}<\epsilon<\omega_{D}$ to 
$-\omega_{D}+\epsilon_{\mu}<\epsilon<\omega_{D}+\epsilon_{\mu}$ 
(where $\epsilon_{\mu}\equiv {\hbar^{2}\mu^{2}\over 2mr^{2}}$):
\begin{equation}
\Delta_{0}=-V\sum_{m}{\Delta_{0}\over 2\sqrt{(\epsilon_{m}+\epsilon_{\mu})^{2}+\Delta_{0}^{2}}}.
\end{equation}
Use of $\epsilon_{m-\mu}+\epsilon_{-m-\mu}=2(\epsilon_{m}+\epsilon_{\mu})$ has been made.
For the $T_{c}$, Eq. (8) shows indeed a minor decrease in $T_{c}$ (for nanoscale systems)
due to the flux, which is {\sl inconsistent} with the ($T=0K$) gap equation, Eq. (7), and has been elusive.$^{5,6}$
However, this decrease is completely negligible for bulk systems and it
originates from the exclusion of the Cooper pairs with negative excitation energy.$^{17,18}$
Since analytic solution is not available, we rely on the numerical calculations and merely emphasize that 
the correction is in the  denominator:
\begin{eqnarray}
1&=& -V\sum_{m}{tanh{\epsilon_{m-\mu}\over 2T_{c}}+tanh{\epsilon_{-m-\mu}\over 2T_{c}}\over 2(\epsilon_{m}+\epsilon_{\mu})}\nonumber\\
&=& -V\sum_{m}{1\over 2(\epsilon_{m}+\epsilon_{\mu})}{sinh{\epsilon_{m}+\epsilon_{\mu}\over T_{c}}\over cosh{\epsilon_{m-\mu}\over 2T_{c}}cosh{\epsilon_{-m-\mu}\over 2T_{c}}}.
\end{eqnarray}
Note that for bulk systems m is of the order of 1,000 near $E_{F}$, while $\mu$ is of the order of 1.

For example, we solve Eqs. (7) and (8) numerically for a superconducting 
cylinder with the inner radius $r_{a}=7,200\AA$, the outer radius 
$r_{b}=8,000\AA$, and the height $L=500\AA$.  
For a thin cylinder, i.e. $r_{b}-r_{a}<<r_{b}$, the electron energy is approximately written as
$\epsilon_{m-\mu}\cong {\hbar^{2}\over 2m}[{(m-\mu)^{2}\over r_{a}r_{b}}+({\ell \pi\over r_{b}-r_{a}})^{2}+({n_{z}\pi\over L})^{2}]$.$^{19,20}$
(More calculational details will be explained in the next section.)
We assumed $E_{F}=1.0eV$ and $\omega_{D}=0.02eV$. 
There are 1,286,200 states in the BCS cutoff range.
Figure 1 shows $\Delta_{0}$ and $T_{c}$ versus $\mu=e\Phi/hc$. 
As can be seen, there is no change in $\Delta_{0}$ (i.e., the condensation 
energy) and $T_{c}$ (i.e., the free energy of superconducting state) due to
the variation of the kinetic energy with the flux $\Phi$. 
For a certain V ($=17.0/\rm{volume}$ [eV]), the gap parameters $\Delta_{0}$ are the same 2.144 886 meV 
for  $\mu=0$ and $\mu=0.25$, whereas the $T_{c}'s$ are 1.212 888 meV ($\mu=0$) and
1.211 921 meV ($\mu=0.25$), respectively. It is noteworthy that the decrease in $T_{c}$
scales approximately $\sim 1/\sqrt{N}$,  with N= number of states in the BCS cutoff range.  Consequently, the  
$T_{c}$ decrease due to the exclusion of the Cooper pairs with negative excitation energy$^{17,18}$
can be completely ignored in the bulk limit.
Note also that $2\Delta_{0}/T_{c}\cong 3.537$, in agreement with the bulk result.

\section{\bf Flux-dependent pairing interaction} 

Now we consider flux-dependence of the pairing matrix element and the
pairing interaction in a superconducting cylinder whose axis coincides
with the z-axis. The inner and outer radii are $r_{a}$ and $r_{b}$, 
and the height is L. First, we determine one-particle eigenfunctions and eigenvalues
in the presence of the Aharonov-Bohm flux $\Phi$.
The Hamiltonian is given by 
\begin{equation}
H_{0}=-{\hbar^{2}\over 2m_{e}}[{\partial^{2}\over \partial r^{2}} +{1\over r}{\partial\over \partial r}-{1\over  r^{2}}(-i{\partial\over \partial \phi}-\mu)^{2}+{\partial^{2}\over \partial z^{2}}]
\end{equation}
where $\mu=e\Phi/hc$.
Assuming a solution of the form
\begin{equation}
\phi(r,\phi,z)=R(r)sin({n_{z}\pi z\over L})e^{im\phi}.
\end{equation}
we obtain the Bessel equation for $R(r)$:
\begin{equation}
R''+{R'\over r}+[k^{2}-{(m-\mu)^{2}\over r^{2}}]R=0.
\end{equation}
From the boundary condition $R(r=r_{a})=R(r=r_{b})=0$, we find the 
radial solutions and eigenvalues:$^{19,20}$
\begin{equation}
R_{m-\mu}(r)={\cal N}[N_{m-\mu}(kr_{a})J_{m-\mu}(kr)-J_{m-\mu}(kr_{a})N_{m-\mu}(kr)]
\end{equation}
\begin{equation}
E_{lmn_{z}}={\hbar^{2}\over 2m_{e}}[k_{ml}^{2}+({n_{z}\pi\over L})^{2}] \ (n_{z}=1,2,3,\cdots,\ m=0,\pm 1,\pm 2,\cdots)
\end{equation}
where $\cal N$ is the normalization constant and $k_{ml}$ is the $l$th root of 
the transcendental equation
\begin{equation}
N_{m-\mu}(kr_{a})J_{m-\mu}(kr_{b})-J_{m-\mu}(kr_{a})N_{m-\mu}(kr_{b})=0.
\end{equation}

From Eq. (16) we calculate the single-particle energy as a function of the flux $\Phi$, shown in Figure 2,
near $E_{F}=1.0$ eV for $r_{a}=75\AA$ and $r_{b}=150\AA$. 
Ignoring the z coordinate, we find 15 basis pair states within the BCS cutoff range for $\omega_{D}=0.02eV$.
We consider a small size ring for simplicity and because of the difficulty in calculating the high order Bessel functions.
As $\mu$ increases, some states near 1.02 eV get out of the BCS cutoff range,
while some states below 0.98 eV are coming in the range. We disregard this complication, 
since this effect will be negligible in the bulk limit. Moreover, it is caused by the 
simple approximation, which can be avoided in the strong-coupling theory.$^{21}$ 
Near $\mu=0.5$ we chose $E_{F}\cong 0.99$ eV, which is obtained from the $E_{F}$ for $\mu=0$ by applying the flux.
Although the spectra near $\mu=0$ and $\mu=0.5$ are similar, they are not exactly the same due to the finite size effect.
Notice the crossings of energy levels at $\Phi=hc/2e\times n$ (with integer n).

Second, we derive the pairing matrix element for an Einstein phonon model
with the phonon Green's function $D(x-x')$. 
Using the equivalent electron-electron potential in the electron-phonon 
problem,$^{11,22,23}$
\begin{equation}
V(x-x')\rightarrow V_{0}D(x-x'),
\end{equation}
with $x=({\bf r},t)$, we get the pairing matrix element: 
\begin{equation}
V_{m,m'}=V_{0}\int_{r_{a}}^{r_{b}}R^{*}_{m'-\mu}(r)R_{-m'-\mu}^{*}(r)R_{-m-\mu}(r)R_{m-\mu}(r)rdr;\ -{1\over 2}+\delta<\mu<{1\over 2}-\delta
\end{equation}
\begin{equation}
V_{m,m'}=V_{0}\int_{r_{a}}^{r_{b}}R^{*}_{m'-\mu}(r)R_{-m'+1-\mu}^{*}(r)R_{-m+1-\mu}(r)R_{m-\mu}(r)rdr;\ {1\over 2}-\delta<\mu<{1}-\delta
\end{equation}
\begin{equation}
V_{m,m'}=V_{0}\int_{r_{a}}^{r_{b}}R^{*}_{m'+1-\mu}(r)R_{-m'+1-\mu}^{*}(r)R_{-m+1-\mu}(r)R_{m+1-\mu}(r)rdr;\ {1}-\delta<\mu<{3\over 2}-\delta
\end{equation}
where $\delta\sim 1/4$.
We have made use of the BCS wavefunction with the flux:
\begin{equation}
\tilde{\phi}_{BCS}=\prod_{m}(u_{m}+v_{m}a^{+}_{m-\mu\uparrow}a^{+}_{-m-\mu\downarrow}|0>; \ -{1\over 2}+\delta<\mu<{1\over 2}-\delta
\end{equation}
\begin{equation}
\tilde{\phi}_{BCS}=\prod_{m}(u_{m}+v_{m}a^{+}_{m-\mu\uparrow}a^{+}_{-m+1-\mu\downarrow}|0>; \ {1\over 2}-\delta<\mu<{1}-\delta
\end{equation}
\begin{equation}
\tilde{\phi}_{BCS}=\prod_{m}(u_{m}+v_{m}a^{+}_{m-\mu\uparrow}a^{+}_{-m+2-\mu\downarrow}|0>; \ {1}-\delta<\mu<{3\over 2}-\delta.
\end{equation}

Finally, we obtain the BCS gap equation$^{14-16}$  
\begin{equation}
\Delta_{m}=-\sum_{m'}V_{mm'}{\Delta_{m'}\over 4\sqrt{({\epsilon_{m'-\mu}+\epsilon_{-m'-\mu}\over 2})^{2}+\Delta_{m'}^{2}}}
(tanh{E_{m'-\mu}\over 2T}+tanh{E_{-m'-\mu}\over 2T})
\end{equation}
where the excitation energy $E_{m-\mu}$ is
\begin{equation}
E_{m-\mu}={1\over 2}(\epsilon_{m-\mu}-\epsilon_{-m-\mu})+
\sqrt{({\epsilon_{m-\mu}+\epsilon_{-m-\mu}\over 2})^{2}+\Delta_{m}^{2}}.
\end{equation}
Since the analytic expression of the matrix element is not available, we
rely on the numerical method to calculate $T_{c}$ from Eq. (24).
Note that Eq. (24) leads to the negative excitation energy, when 
$\epsilon_{m-\mu}$ and $\epsilon_{-m-\mu}$ have different signs near $E_{F}$,
as in the cases of a superconductor with a uniform exchange field$^{17}$ and
the current-carrying superconducting state.$^{18}$
Nevertheless, in this case the possibility of the negative excitation energy is not significant as shown in
Fig. 1, because $ m \sim 1,000$ near $E_{F}$ while $\mu \sim 1 $  in the bulk
limit.  In other words, the exclusion of the contribution of the negative excitation
does not lead to any noticeable effect in $T_{c}$, as shown in Fig. 1. 

\section{\bf Explanation of Flux Quantization} 

Flux quantization may be explained from the flux-dependence of the
pairing matrix element $V_{mm'}$.
We stress the periodicity of the matrix element with the flux:
\begin{equation}
V_{mm'}(\Phi={hc\over 2e}\times \rm{even\ n})=V_{mm'}(\Phi=0).
\end{equation}
\begin{equation}
V_{mm'}(\Phi={hc\over 2e}\times \rm{odd\ n})\cong V_{mm'}(\Phi=0).
\end{equation}
This behavior is caused by the energy level crossings at 
 $\Phi=hc/2e\times n$ (or $\mu=\rm{integer\ or\ half-integer}$).
It was shown that the level crossings (at integer $\mu$) are due to the
cylindrical symmetry of the system.$^{4,20}$
Since for odd n the energy level crossing occurs at lower energy 
than for even n, the matrix elements are slightly different for
even and odd n's.

In the neighborhood of $\Phi=0+$ we pair states ($m-\mu\uparrow$)
and ($-m-\mu\downarrow$) which are not degenerate and have a different 
density distribution.$^{10,11}$ Accordingly, the pairing matrix element 
$V_{mm'}$ is decreasing with the flux, compared to $V_{mm'}$ at zero flux.
For a large cylinder, we may employ the asymptotic radial wavefunction:$^{24}$ 
\begin{equation}
R_{m-\mu}(r)={\cal N}\{[1+a_{m-\mu}{1\over r_{a}r}-b_{m-\mu}({1\over 2r^{2}}+{1\over 2r_{a}^{2}})] {sin k_{l}(r-r_{a})\over \sqrt{r}}\}
\end{equation}
where
\begin{equation}
a_{m-\mu}= [{4(m-\mu)^{2}-1\over 8k_{l}}]^{2}, \ b_{m-\mu}= {[4(m-\mu)^{2}-1][4(m-\mu)^{2}-9]\over (8k_{l})^{2}},
\end{equation}
and
\begin{equation}
k_{l}={l\pi\over r_{b}-r_{a}}.
\end{equation}
It is obvious that the algebraic terms lead to the different density 
distributions of states $(m-\mu\uparrow)$ and $(-m-\mu\downarrow)$ and the
decrease of the matrix element:
\begin{equation}
V_{mm'}(\Phi)\approx [1-C'({2m^{3}\over k_{l}^{2}})^{2}\mu^{2}] [1-C'({2m'^{3}\over k_{l'}^{2}})^{2}\mu^{2}]\times V_{mm'}(\Phi=0),
\end{equation}
where $C'\cong (1/r_{a}r_{b}-1/2r_{b}^{2}-1/2r_{a}^{2})^{2}$.

Figure 3 shows the radial density distributions $|R(r)|^{2}r$ of the $m=\pm 54$ states when $\mu=0.25$, (i.e., 
$m-\mu=53.75, -m-\mu=-54.25$) for the same ring as Fig. 2.
The solid line corresponds to $m-\mu=54.75$, whereas the dotted line corresponds
to $-m-\mu=-54.25$, in which the electron is rotating faster, due to bigger angular momentum projection along the z-axis, leading to the shift of the radial distribution towards the outer radius.  
The difference in the radial density distributions then induces the decrease of
the pairing matrix element with increasing $\mu$. Figure 4 shows the matrix
element $V_{mm'}\equiv V_{ij}$ for $\mu=0$. Notice that it is $15\times 15$ symmetric matrix, 
since there are 15 pair states in the BCS cutoff range. The decrease of the matrix 
element with $\mu$ is not big enough to be noticeable in Fig. 4, though. 
We just mention the 
decrease of biggest matrix element $V_{m=53,m'=51}\equiv V_{i=2,j=3}$, 
between pair states with the same $l$ value, from 0.0002450 to 0.0002448.

This observation and the periodicity of $V_{mm'}\equiv V_{i,j}$ prove that flux
quantization is due to the minimum of the pairing energy of the 
superconducting cylinders as a function of the flux. 
Accordingly, the free energy of the superconducting state also has minimum 
at $\Phi=hc/2e\times n$ since the contribution of the kinetic energy is 
constant. 
In other words, pairing between the degenerate states caused by the energy 
level crossing at $\Phi=hc/2e\times n$ is the physical origin of flux 
quantization. Thus the Little-Parks oscillations follow accordingly.
From Eqs. (24) and (31) we 
may estimate roughly the decreases of the energy gap at $T=0K$ and $T_{c}$:
\begin{eqnarray}
\Delta_{m}(\Phi)& \sim& \Delta_{m}(\Phi=0)(1-8C'<{m^{6}\over k_{l}^{4}}>\mu^{2}/N_{0}V_{0}),\\
T_{c} &\sim& T_{c0} (1-8C'<{m^{6}\over k_{l}^{4}}>\mu^{2}/N_{0}V_{0}),
\end{eqnarray}
where $N_{0}$ is the density of states at $E_{F}$ and $< >$ denotes the average 
over the states in the BCS cutoff range. $T_{c0}$ is the transition temperature in the absence of the magnetic flux.

Figure 5 shows the $T=0K$ {\sl anisotropic} gap parameter $\Delta_{i}$ for $\mu=0$, calculated 
from Eq. (24) by employing the matrix element $V_{i,j}\equiv V_{mm'}$ as in Fig. 4. 
Even though the coupling between the second and third
states is strongest, we find the biggest gap at the tenth state, i.e., 
$\Delta_{i=10}$, due to its strong couplings to several neighbor states.

Figure 6 displays the decrease of the 
(average) energy gap with $\mu$. The solid line denotes the ($T=0K$) average gap parameter, 
$\bar{\Delta}=\sum_{i}\Delta_{i}/15$.
It should be noticed that this decrease, so much enhanced as a result of the 
finite size effect, persists even in the bulk limit, as Eq. (32) shows.
The different variations near $\mu=0$ and $\mu=0.5$ are also due to the finite size effect. The dashed line shows the isotropic
gap parameter, according to the assumptions of Byers and Yang.$^{4}$
It exhibits a small decrease with the flux too, which quickly disappears with
the increase of the system size. For the isotropic solution, we included 
$V_{ii}=V$ besides $V_{ij}=V=$ constant. Therefore, we get the bigger gap.
The semicircle is the $T_{c}$ when $\mu=0$.
Since, as the flux is increased, the finite temperature gap equation, 
Eq. (24) eliminates the contributions from the Cooper pairs with negative
excitation energy,$^{17,18}$ $T_{c}$ drops to zero immediately. 
If we consider the Cooper problem,$^{25}$ it is easy to understand why we need 
to eliminate the Cooper pair basis states with different signs in energy near 
$E_{F}$.  However, this behavior is greatly amplified because of the finite size effect and it will 
not cause any problem in realistic systems, 
as shown in Fig. 1. Furthermore, the strong coupling theory$^{21}$ will reduce the 
effect even more.  For nanoscale systems it may be desirable to use 
the canonical ensemble approach.$^{26}$ 
We expect that for a reasonably large system, we find a continuous (small) variation of the $T_{c}$, consistent with 
the variation of the energy gap. 

\section{\bf Discussion } 

It is clear that more study is needed to understand the details of flux 
quantization. In particular, it is essential to consider a big hollow cylinder
for comparison with the experiment, which requires high order Bessel functions. 
This study indicates that the parabolic background$^{5,6}$ in the 
$T_{c}$ oscillations may be due to the magnetic field penetration into the 
superconductor, which is not addressed here.
It is also interesting that disordered $Au_{0.7}In_{0.3}$ cylinders$^{27}$ show 
half-integer flux quantization as in the case of high $T_{c}$ cuprates.$^{3}$ 

This study also implies that the exact calculation of the 
eigenenergies using high order Bessel functions and the magnetic field penetration 
into the metals are important in calculating the persistent current in normal metals.

Finally, flux quantization plays an important role in the vortex state$^{28}$ which
may be understood by generalizing this approach.

\section{\bf Conclusion } 
We have shown that flux quantization and the Little-Parks oscillations in 
superconducting cylinders are
due to the flux dependence of the pairing interaction, while the flux
dependence of the kinetic energy is canceled as evidenced by the observation 
of the persistent current in normal metals. 
When the flux $\Phi$ is given by $\Phi=hc/2e\times n$ (with integer n), 
the pairing interaction and the free energy are unchanged (even n) or almost unchanged (odd n), 
due to the degenerate-state pairing,  
compared to those in the absence of the flux. Accordingly, flux quantization 
and the Little-Parks oscillations are obtained.

\vspace{2pc}

\centerline{\bf ACKNOWLEDGMENTS}

M.P. thanks NSF-EPSCOR (Grant No. EPS9874782) for financial support.
YJK is grateful to Faculty of Arts and Sciences at UPR-Mayaguez for release
time.

\vfill\eject

\begin{figure}
\caption{Energy gap and $T_{c}$ as a function of the flux $\Phi$ for a cylinder with $r_{a}=7,200\AA$, $r_{b}=8,000\AA$, and $L=500\AA$, following Byers and Yang. Energy gap does not change at all, while $T_{c}$ shows negligible oscillations.} 
\end{figure}

\begin{figure}
\caption{One-particle energy levels vs $\mu=e\Phi/hc$ near the Fermi energy ($E_{F}=1.0 eV$ for $\mu=0$ and $E_{F}\simeq 0.99 eV$ for $\mu=0.5$) for a ring with $r_{a}=75\AA$ and $r_{b}=150\AA$. Note the level crossings at $\mu=0$ and 0.5.} 
\end{figure}

\begin{figure}
\caption{Radial probability density $|R(r)|^{2}r$ of states $m-\mu=53.75$ (solid line) and $-m-\mu=-54.25$ (dotted line) for the same ring as Fig. 2. Notice the shift of the radial distribution of $-m-\mu=-54.25$ state to the outer radius. } 
\end{figure}

\begin{figure}
\caption{Pairing matrix element $V_{mm'}\equiv V_{ij}$ for the same ring as Figs. 2 and 3. The biggest element $V_{i=2,j=3}\equiv V_{m=53,m'=51}$ corresponds to the pair states with the same $l$ value.  The diagonal term $V_{mm}\equiv V_{ii}$ does not contribute to superconductivity. } 
\end{figure}

\begin{figure}
\caption{Anisotropic gap parameter $\Delta_{i}$ for $\mu=0$ at $T=0K$, corresponding to the matrix element, $V_{ij}$, as in Fig. 4.  } 
\end{figure}

\begin{figure}
\caption{Average energy gap parameter (solid line), $\bar{\Delta}=\sum_{i}\Delta_{i}/15$, (at $T=0K$) vs $\mu$ for the same ring as Figs. 2-5. The dashed line shows the isotropic gap parameter vs $\mu$ following Byers and Yang. The semicircle denotes the $T_{c}$ for $\mu=0$, corresponding to the anisotropic gap parameter, $\Delta_{i}$.}
\end{figure}

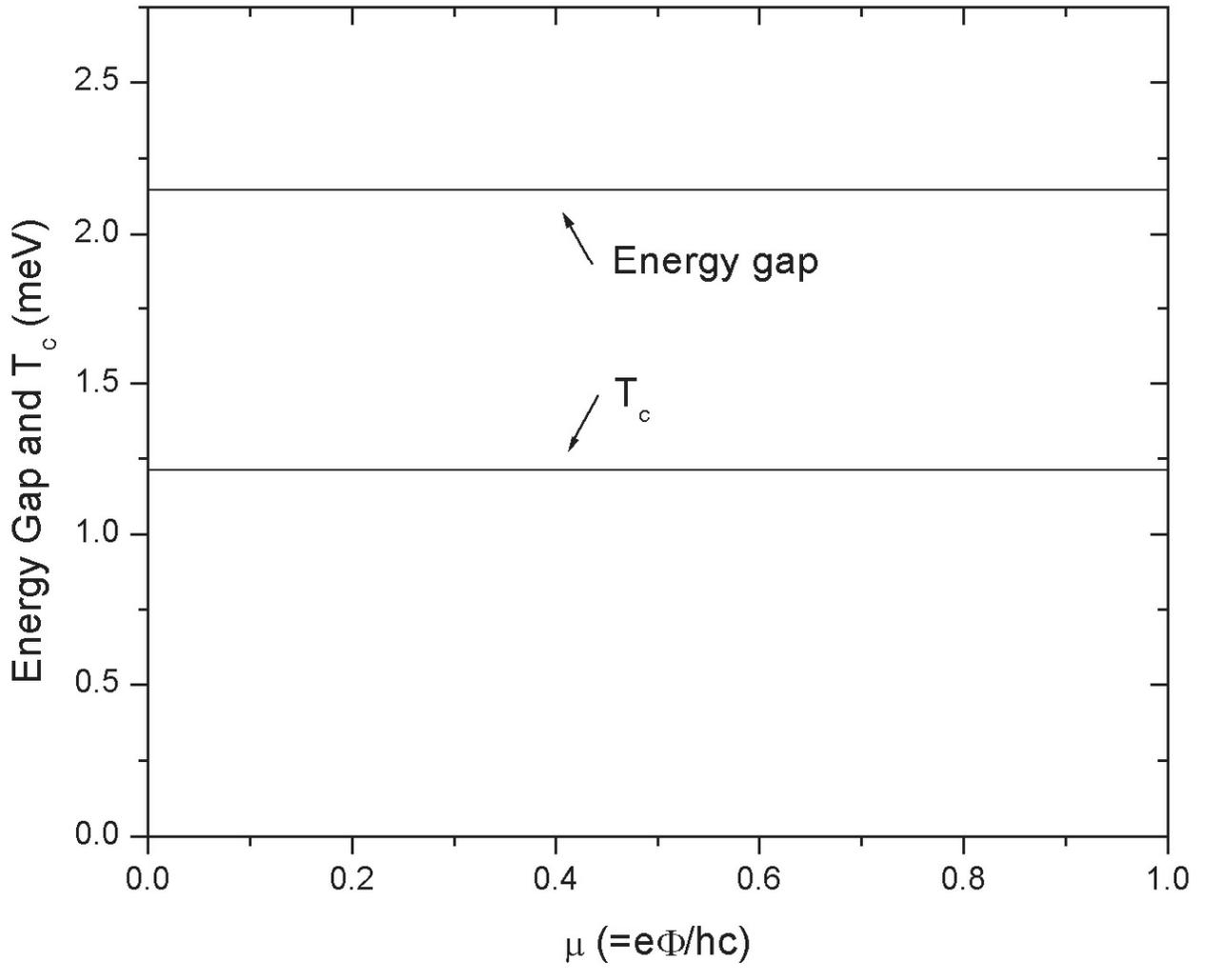
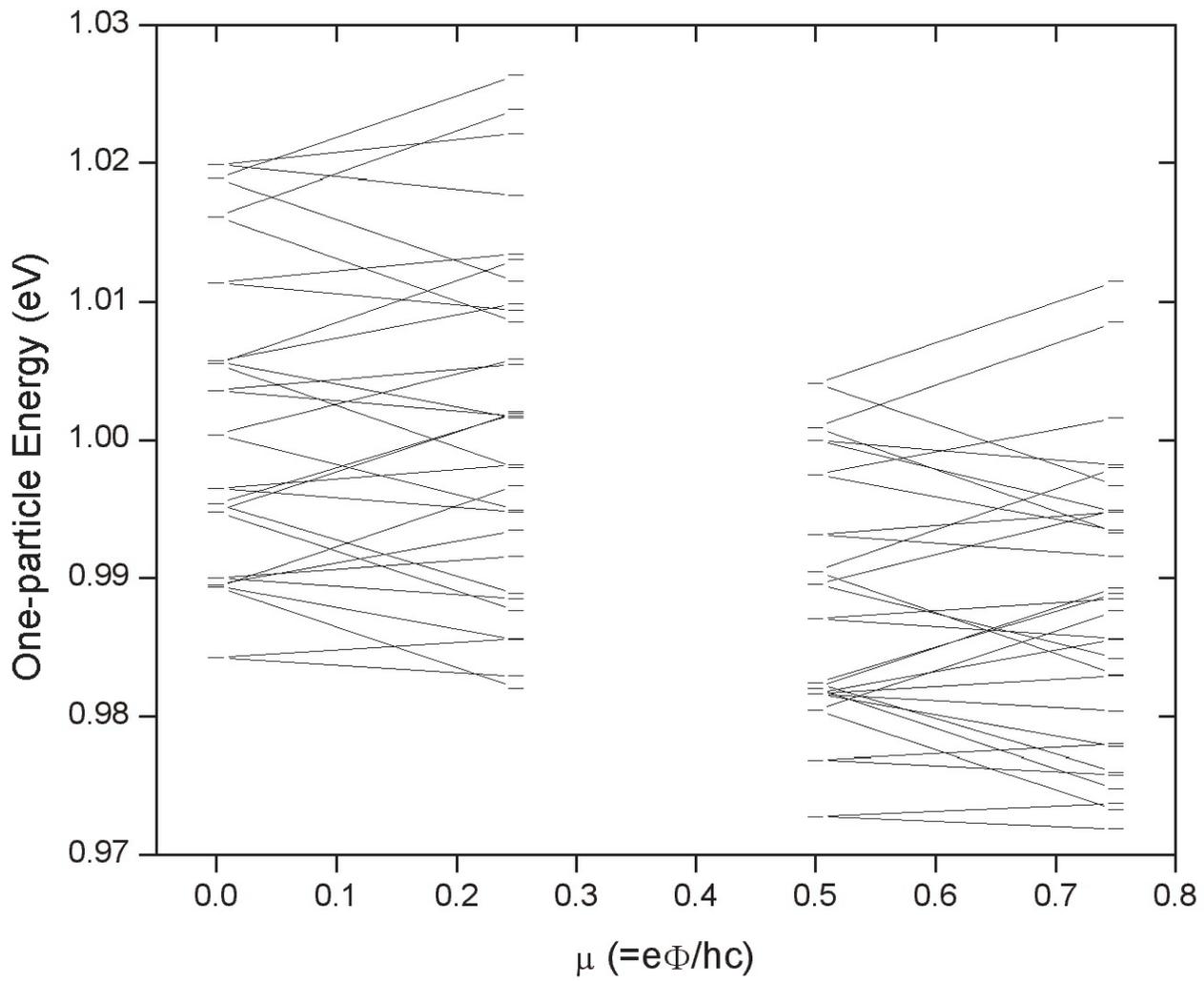
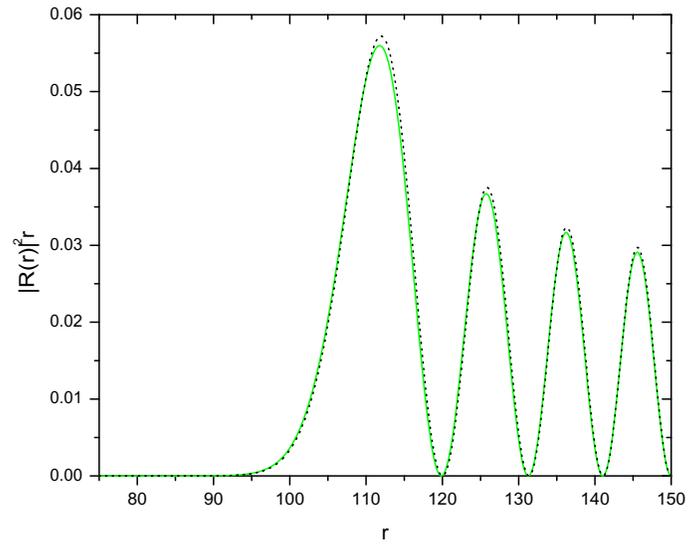
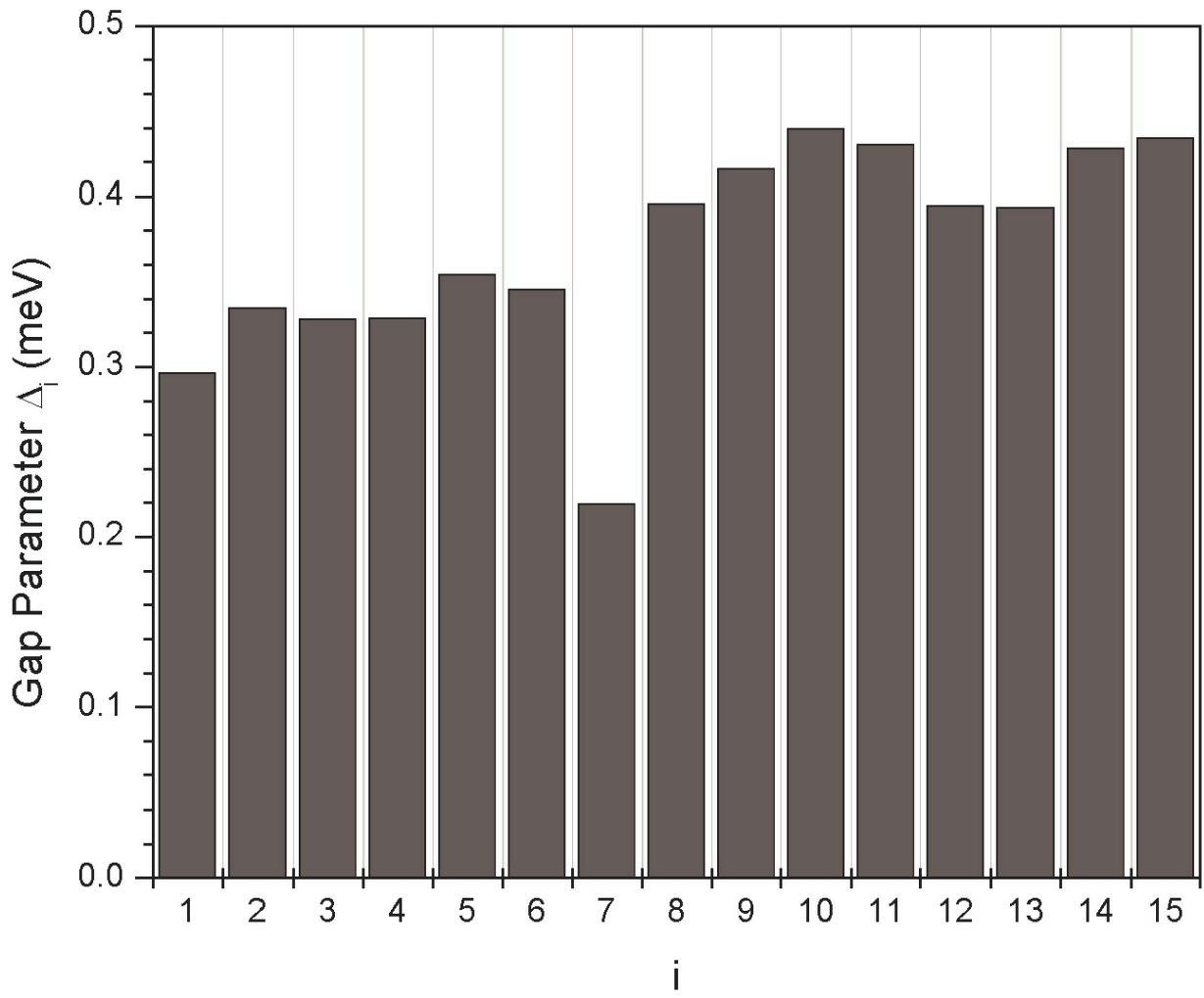
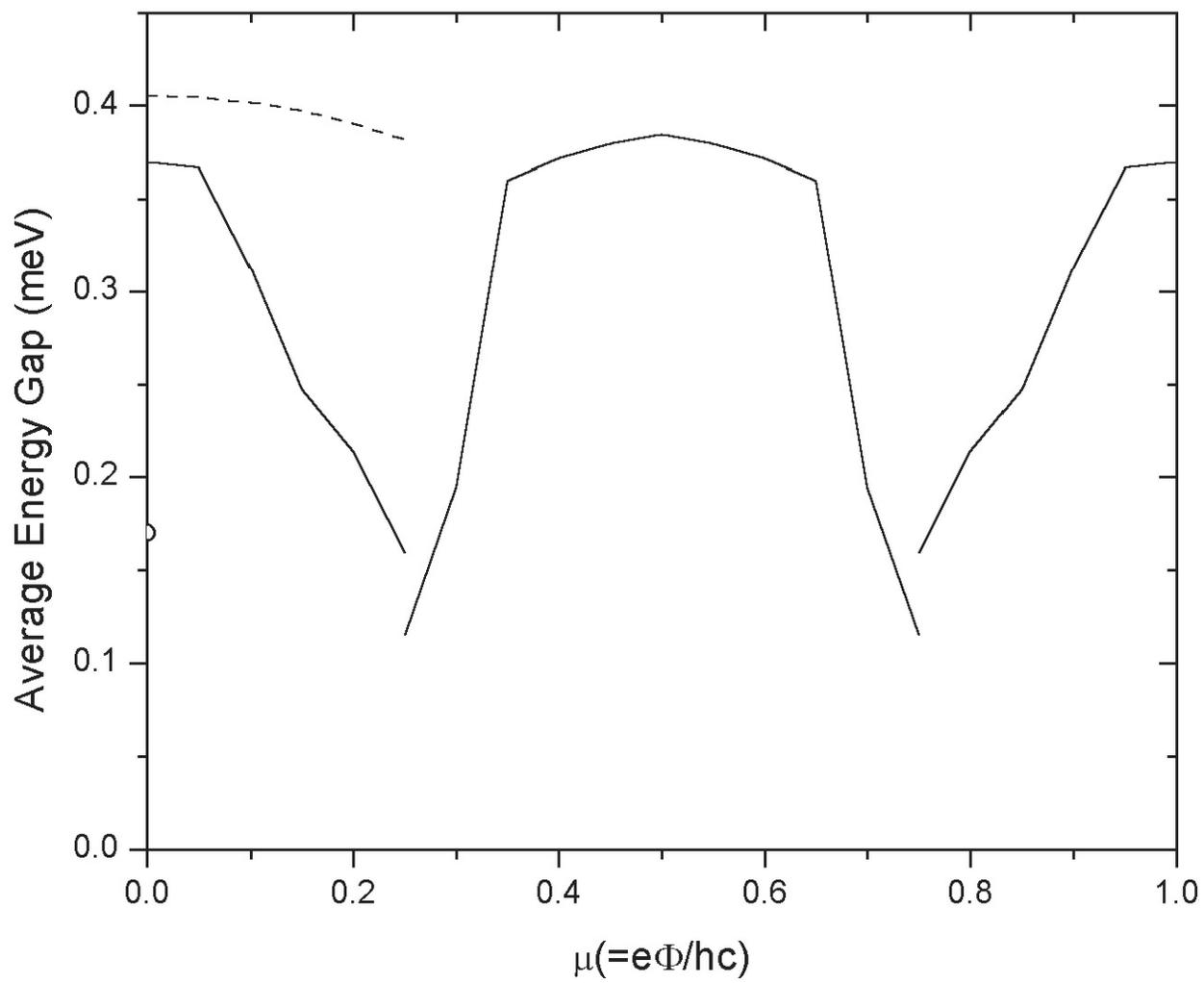
\end{document}